\documentclass[10pt,conference]{IEEEtran}

\AtBeginDocument{%
 \providecommand\BibTeX{{%
   \normalfont B\kern-0.5em{\scshape i\kern-0.25em b}\kern-0.8em\TeX}}}
    
\usepackage{xspace}

\usepackage{array}
\newcolumntype{L}{>{\arraybackslash}m{16cm}}
\newcolumntype{C}[1]{>{\centering\let\newline\\arraybackslash\hspace{0pt}}m{#1}}
\newcolumntype{R}[1]{>{\raggedleft\let\newline\\arraybackslash\hspace{0pt}}m{#1}}
\usepackage[numbered]{bookmark}
\usepackage{tcolorbox}
\usepackage{graphicx}
\usepackage{xcolor}
\usepackage{stfloats}
\usepackage[justification=centering]{caption}
\usepackage{lineno}
\usepackage{dirtytalk}
\usepackage{amsmath}
\usepackage{pgfplots, pgfplotstable}
\usetikzlibrary{patterns}

\usepackage{titlesec}
\titleformat{\subsubsection}[runin] 
  {\normalfont\bfseries} 
  {\thesubsubsection} 
  {1em} 
  {} 
  [\newline] 
  
\usepackage[english]{babel}
\usepackage{textcomp}
\usepackage{mathtools}
\usepackage{ltxtable}
\usepackage{algorithm}
\usepackage{algpseudocode}
\usepackage{textcomp}

\usepackage{tcolorbox}
\usepackage{multirow}
\usepackage{graphicx}
\usepackage{pgf-pie}
\definecolor{wedge1}{RGB}{ 190  30  46}
\definecolor{wedge2}{RGB}{ 240  65  54}
\definecolor{wedge3}{RGB}{ 241  90  43}
\definecolor{wedge4}{RGB}{ 247 148  30}
\definecolor{wedge5}{RGB}{  43  56 144}
\definecolor{wedge6}{RGB}{  28 117 188}
\definecolor{wedge7}{RGB}{  40 170 225}
\definecolor{wedge8}{RGB}{ 119 179 225}
\definecolor{wedge9}{RGB}{ 181 212 239}
\definecolor{wedge10}{RGB}{  0 104  56}
\definecolor{wedge11}{RGB}{  0 148  69}
\definecolor{wedge12}{RGB}{ 57 181  74}
\definecolor{wedge13}{RGB}{141 199  63}
\definecolor{wedge14}{RGB}{215 244  34}
\definecolor{wedge15}{RGB}{249 237  50}
\definecolor{wedge16}{RGB}{248 241 148}
\definecolor{wedge17}{RGB}{242 245 205}
\definecolor{wedge18}{RGB}{123  82  49}
\definecolor{wedge19}{RGB}{104  73 158}
\definecolor{wedge20}{RGB}{102  45 145}
\definecolor{wedge21}{RGB}{148 149 151}
\definecolor{wedge22}{RGB}{ 204 50 153}
\definecolor{wedge23}{RGB}{ 79 47 79}
\definecolor{wedge24}{RGB}{ 173 234 234}
\definecolor{wedge25}{RGB}{ 216 191 216}
\definecolor{wedge26}{RGB}{  43  56 144}
\definecolor{wedge27}{RGB}{  40 170 225}
\definecolor{wedge28}{RGB}{ 119 179 225}
\definecolor{wedge29}{RGB}{ 181 212 239}
\definecolor{wedge30}{RGB}{  0 104  56}
\definecolor{wedge31}{RGB}{  0 148  69}
\definecolor{wedge32}{RGB}{ 57 181  74}

\usetikzlibrary{chains}

\pgfmathsetmacro\startAngle{90-3.6/2}
\pgfmathsetmacro\radius{+5}
\pgfmathsetmacro\maxLeg{+12}
\pgfmathsetmacro\legBound{+60}
\pgfmathsetmacro\legSpacing{2*\legBound/(\maxLeg-1)}

\usepackage{verbatim}
\pgfplotsset{compat=1.14}
\usetikzlibrary{patterns,}
\usepackage{hyperref}
\usepackage{hyperxmp}
\usepackage{graphicx}
\usepackage{tabularx,booktabs}
\usepackage{array}
\usepackage{bbding}
\usepackage{pifont}
\usepackage{wasysym}
\usepackage{caption}
\usepackage{dirtytalk}
\usepackage{subcaption}
\usepackage{tabularx}
\usepackage{sfmath}
\usepackage{array, booktabs, makecell}
\usepackage{color}
\usepackage{csquotes}
\usepackage{url}
\usepackage{comment}
\usepackage{tikz}
\usepackage{balance}
\usepackage{listings}
\usepackage{booktabs} 
\usepackage{soul} 
\usetikzlibrary{fit} 
\usetikzlibrary{positioning}
\usetikzlibrary{arrows}
\usetikzlibrary{shapes.multipart}
\usepackage{adjustbox}
\usepackage{float}
\usepackage{rotating}
\usepackage{nth}
\DeclareCaptionType{TextBox}
\usepackage{pstricks}
\usepackage{placeins}
\usepackage{afterpage}
\usepackage{multirow} 
\usepackage{textcomp}
\usepackage{multicol}
\usepackage{varwidth} 

\begin{document}

\title{\huge Exploring Prompt Patterns in AI-Assisted Code Generation: Towards Faster and More Effective Developer-AI Collaboration}

\author{
\IEEEauthorblockN{ 
Sophia DiCuffa\IEEEauthorrefmark{1},
Amanda Zambrana\IEEEauthorrefmark{1},
Priyanshi Yadav\IEEEauthorrefmark{1},
Sashidhar Madiraju,
Khushi Suman,
Eman Abdullah AlOmar}
\IEEEauthorblockA{Stevens Institute of Technology, Hoboken, New Jersey, USA\\
\{sdicuffa,azambran,pyadav6,smadiraj1,ksuman,ealomar\}@stevens.edu\\
}} 
\maketitle

\begingroup
\renewcommand\thefootnote{\IEEEauthorrefmark{1}}
\footnotetext{These authors contributed equally to this work.}
\endgroup

\begin{abstract}
The growing integration of AI tools in software development, particularly Large Language Models (LLMs) such as ChatGPT, has revolutionized how developers approach coding tasks. However, achieving high-quality code often requires iterative interactions, which can be time-consuming and inefficient. This paper explores the application of structured prompt patterns to minimize the number of interactions required for satisfactory AI-assisted code generation. Using the DevGPT dataset, we analyzed seven distinct prompt patterns to evaluate their effectiveness in reducing back-and-forth communication between developers and AI. Our findings highlight patterns such as ``Context and Instruction" and ``Recipe" as particularly effective in achieving high-quality outputs with minimal iterations. The study emphasizes the potential for prompt engineering to streamline developer-AI collaboration, providing practical insights into crafting prompts that balance precision, efficiency, and clarity.
\end{abstract}



\begin{IEEEkeywords}
prompt patterns, DevGPT, LLMs
\end{IEEEkeywords}


\section{Introduction}
\label{Section:Introduction}
The growing popularity of AI-based code generation and assistance tools has revolutionized software development by assisting in tasks ranging from code writing to debugging \cite{alomar2024refactor,chouchen2024software,depalma2024exploring}. However, it often takes long, iterative conversations between developers and AI-based coding assistants to achieve high-quality software outputs. While some developers find iterative development to be helpful, others find it a quite tedious and time-consuming task. In our study, we suggest that developers begin their AI-assisted code development with proper prompt engineering in order to initially establish a means for creating high-quality code without spending excessive amounts of time on iterative conversations with their AI assistants.

The application of prompt engineering, the process of designing inputs to guide AI models to produce desired outputs, can help reduce the time it takes to produce high-quality software with an AI assistant. This paper examines how prompt engineering, specifically through the use of prompt patterns, can minimize interactions with Large Language Models (LLMs) in order to effectively and efficiently elicit the desired output. Prompt patterns are structured methods for creating reusable prompts that guide interactions with LLMs. They help to improve the accuracy of the results generated by LLMs like ChatGPT. 

In our experiments, we analyze a dataset called DevGPT \cite{devGPT}, consisting of conversations between developers and the ChatGPT AI assistant, screening for a set of seven prompt patterns to classify which patterns have been used in each conversation. Then, by analyzing conversations between human developers and AI assistants within the DevGPT dataset, the number of prompts used in each session is quantified, with the total number of interactions indicating how many iterations it took for the developer to reach the desired output. The findings are intended to provide insights into which prompt patterns developers should use to enhance the effectiveness of their collaboration with AI to produce high-quality code in fewer interactions. 

\section{Problem Statement}
\label{Section:Problem}
The increasing reliance on AI-assisted code generation tools, such as OpenAI’s ChatGPT, has transformed software development. However, it often takes long, iterative conversations between developers and AI-based coding assistants to achieve high-quality software outputs. Human developers find it undesirable to conduct such conversations with AI assistants, finding it a waste of valuable time. 

In order to minimize the number of iterations of a conversation with an AI assistant required to achieve the desired output, developers need to create direct prompts that are clear and specific to provide precise guidance for the AI. Unclear instructions frequently result in code issues like unnecessary complexity, bugs, or code duplication. To better solve this problem, we suggest that developers apply prompt patterns to help reduce these issues and achieve high-quality software collaboration with AI assistants promptly.  By applying prompt patterns rather than simply asking loose and unspecific questions, developers can reduce the number of iterations necessary to elicit a satisfactory response from ChatGPT. 

Our study aims at determining the best-fit  prompt patterns to apply so that the overall number of interactions can be minimized when using ChatGPT for the desired output code. We analyze conversations between developers and ChatGPT as the AI assistant from the DevGPT dataset \cite{devGPT} to detect prompt patterns used within those conversations. This process allows us to determine and suggest specific prompt patterns that prove to be effective for improving AI conversations such that they enable the developer to achieve their desired output in the minimum amount of interactions with the AI.

\section{Related Work}
\label{Section:RelatedWork}
AI-based code generation tools, such as ChatGPT, have revolutionized software development by assisting with tasks like code generation, debugging, and optimization. However, the process of generating code with an AI-assistant has its challenges. For instance, the effectiveness of these tools often hinges on the quality of the prompts provided by developers, as poorly structured or ambiguous prompts can lead to suboptimal outputs or increased interaction time with the AI tool. This section reviews existing research on AI-assisted coding, emphasizing the role of prompt patterns in achieving efficient and high-quality results. It also presents a comparative analysis of studies that influenced our decision of the seven prompt patterns used in our experiments.

Several studies have explored the potential of AI tools, such as OpenAI’s GPT, to influence software development outcomes.  For example, it has been demonstrated that systematically designed prompt templates can significantly improve the quality of responses from large language models (LLMs) in coding tasks \cite{PromptDesign}. Their research highlighted the effectiveness of structured instructions in clarifying task goals and ensuring more reliable outputs, emphasizing the importance of aligning prompts with desired outcomes. Wei et al. introduced the concept of instruction induction to guide LLMs in performing complex, multi-step tasks \cite{COT}. They found that detailed, hierarchical instructions improve the model's performance on reasoning and coding challenges. These underscore the need for deliberate prompt crafting, a concept that is central to our investigation of specific prompt patterns for software development. In our study, we aim to extend this research by analyzing how developers can engineer their prompts for AI assistants such that the desired output and quality is achieved without having to conduct continuous, time-consuming conversations. 

The use of prompt patterns to optimize interactions with AI tools has been an emerging research focus. Reynolds and McDonell proposed that prompt-based fine-tuning methods enable more flexible task adaptation, particularly for domain-specific problems \cite{PPforLLM}. They emphasized that persona-based patterns, where the AI is assigned a specific role or identity, significantly improve contextual understanding. 

The reviewed studies provide a strong foundation for understanding how prompt engineering can optimize developer-AI interactions. However, limitations such as a lack of empirical focus on software-specific applications or proactive prompt strategies suggest opportunities for further exploration from the aforementioned studies. Our study builds on this work by systematically evaluating how specific prompt patterns impact interaction efficiency and output quality, particularly in scenarios requiring minimal iterative communication with an AI assistant. Overall, in our study, we explore how developers can apply prompt engineering to tailor their interactions with AI-based tools to produce better-quality
code through minimal iterative communication, ultimately saving valuable time without sacrificing quality. While the related works do not all focus exclusively on software development, their findings are adaptable to this domain, demonstrating how well-designed prompt patterns can support developers in achieving specific outcomes efficiently.

\section{Study Design}
\label{Section:Methodology}
\subsection{Analysis of ChatGPT Conversations}
In our proposed solution, we suggest that developers apply prompt patterns in order to elicit a satisfactory response from an AI assistant in the least possible amount of interactions. We aim to determine which prompt patterns are the most effective in eliciting such a response in the most efficient way. In order to determine those patterns, we analyze a set of ChatGPT conversations conducted between software developers and ChatGPT as the AI assistant leveraged for coding prompts. By applying a keyword searching algorithm to detect words indicative of specific prompt patterns, we predict which pattern was used in each conversation listed in our dataset. Furthermore, each conversation contains an associated number of interactions, indicating the number of conversation points iterated back and forth between the developer and the AI assistant. We analyze these numbers against the prompt pattern detected within the corresponding conversation in order to pinpoint and suggest the best prompt patterns for developers to use. 

To design our approach to solving the aforementioned problem of the burden of time-consuming interactions with AI assistants necessary for reaching the desired output results, we first determined that we would need to study a dataset of developer conversations with an AI assistant. We wanted to analyze real developer interactions with ChatGPT, leading us to the DevGPT dataset \cite{devGPT}. The DevGPT dataset \cite{devGPT} contains a collection of developer-AI conversations centered around various discussions. The dataset captures detailed exchanges between developers and ChatGPT, providing a valuable resource for studying prompt structures and their effectiveness in producing relevant results. Then, we began designing the approach of evaluating prompt patterns used in order to determine the best-fit patterns to be suggested for use in prompt engineering.

\subsection{Selecting Prompt Patterns}
To predict the specific prompt pattern used by developers in each conversation present in our dataset, we will first outline a set of prompt patterns that we will be screening for in our experiments. In order to determine which prompt patterns to include in our study, we took inspiration from the related works we evaluated earlier in the paper. These studies collectively informed our selection by emphasizing the effectiveness, flexibility, and applicability of these patterns in various natural language processing (NLP) and user interaction scenarios. By drawing insights from the findings of those six related studies, we were able to highlight seven patterns for our study, outlined in Table \ref{tab:promptpatterns}. 

\begin{table*}[]
\centering
\caption{Summary of Prompt Patterns used}
\label{tab:promptpatterns}
\resizebox{\textwidth}{!}{%
\begin{tabular}{|p{4cm}|p{5cm}|p{8cm}|}  
\hline
\textbf{Prompt Pattern} & \textbf{Description} & \textbf{Example from DevGPT Dataset} \\
\hline
Persona Pattern & The user defines a specific role or persona for the AI to assume during the 
interaction to deliver tailored information or services. (also called 
“role-playing” prompts).& “You are an Odoo ERP implentation expert.  The default URL parameters land instead on the ""Description"" tab of the Task form in the Odoo app ""Project"".    Your task is to create a URL that lands a user on the ""Sub-tasks"" tab of the Task form in the Odoo app ""Project"". If there is no specific URL parameters to complete this task, provide some guidance on the appropriate python extension or customization.” (Issue 126140) \cite{devGPT} \\
\hline
Recipe Pattern & The user outlines a step-by-step sequence of instructions provided to the 
AI to achieve a desired result or specific task. Especially useful for getting a 
process outline or sequence of actions from the AI. & “Write a function called readfile(path): it opens that file and reads the first 512 bytes. Then it splits that text on newlines to get just the first to lines, and runs that regular expression against  them to find the encoding.  If the encoding is missing it assumes utf-8. Finally it reads the entire file using the detected encoding and returns it” (Issue 18) \cite{devGPT} \\
\hline
Template Pattern & The user describes a reusable structure or format given to the AI to fill in 
with specific details such that the AI can give its response in that certain 
template, pattern, or structure. & “ [...]Requirements:
- Assign a `precision` label for ALL records
- Respond in CSV format using a pipe (i.e. ""|"") delimiter with the headers `id`, `precision` where `id` is the `id` associated with each record
- Include the headers in the result 
- Respond with ONLY the CSV content, do not include explanation of any kind” (Issue 3) \cite{devGPT}
 \\
\hline
Output Automator Pattern & The user’s prompt emphasizes the structure and rules for generating 
outputs, ensuring the AI produces content in a machine-readable or 
predictable format. & “For iPhone 6+ (4K 30 FPS) I got new data. 7 seconds video uses 40.8MB, 4 seconds video uses 19.5, 3 seconds video uses 19.2 MB. Calculate for iPhone 6+ (4K 30 FPS): how long I should record a video to get 15, 30, 45, 50, 55 and 60 MB video file. Show result in table.” (Issue 2773) \cite{devGPT} \\
\hline
Instructions-Based Pattern & The user inputs a simple and straightforward command or directive on how 
to provide the desired output without additional context or persona 
definitions. & "Help me design some rust code for no-std that supports the following.
High level description
Rotations are a key component of attitude and orientation parameters. At first, ANISE only supports Direct Cosine Matrix math. This is a redundant representation of rotations and therefore not an optimal one.
The purpose of this issue is to design and implement a correct SO(3) group for use in ANISE. ” (Issue 35) \cite{devGPT} \\
\hline
Context and Instructions Pattern & The user defines a mixture of background information and specific 
instructions provided to guide the AI response. & “I am using the package react-native-image-crop-picker to allow the user to select a video from their iOS device. After clicking on the video, the package shows a ""Processing assets..."" string for the duration of time that it takes to select and compress the video. I would like to patch this package so that I can return the percentage of time completed that the image processor will take.” (Issue 1941) \cite{devGPT} \\
\hline
Question Pattern & This is the simplest prompt pattern, where the user asks a question aimed 
at eliciting an informative or helpful response from the AI. No additional 
context or template/style for how to give the desired output is provided. & “How can i make github notifications show up in discord?” (Issue 7) \cite{devGPT} \\
\hline
\end{tabular}%
}
\end{table*}

The following list provides a more specific explanation of how our evaluated related works influenced our set of prompt patterns used: 
\begin{itemize}
    \item \textbf{Persona Pattern} - The persona pattern aligns with findings in  \cite{reynolds2021prompt}, which discusses how creating conversational or task-specific personas improves user engagement and helps models interpret inputs more contextually \cite{PPforLLM}. Similarly, Liu et al.  highlight that domain-specific prompts often improve performance on tasks requiring specialized understanding, which a persona pattern naturally facilitates \cite{PromptDesign}.
    \item \textbf{Recipe Pattern} - Ye et al. illustrate the utility of step-by-step instructions in improving task accuracy and user comprehension \cite{instructionInduction}. The recipe pattern, with its focus on sequential steps, reflects this structure, making it effective for complex or multi-step developer tasks such as debugging or algorithm generation. 
    \item \textbf{Template Pattern} - Previous studies \cite{FewShotLearners,humanLevel} emphasize the importance of templates in standardizing input formats to optimize LLM performance across tasks. This pattern supports efficient reuse of prompts, particularly in scenarios requiring repeated task structures, such as documentation generation or data transformation. 
    \item \textbf{Output Automator Pattern} - This pattern draws from findings in \cite{COT}, which demonstrate how structuring outputs can improve reasoning and logic tasks. Automaton patterns emphasize predefined output formats, which are especially relevant to software developers aiming to automate specific outputs like configurations or scripts.
    \item \textbf{Instruction-Based Pattern} - Previous studies \cite{instructionInduction,humanLevel} underscore the effectiveness of instruction-driven prompts in improving model task adherence and reducing ambiguity. Instructions-based patterns are critical for software developers who often require precise and unambiguous responses from AI. 
    \item \textbf{Context and Instructions Patterns} - Contextual prompting methods discussed in \cite{humanLevel,FewShotLearners} influenced the inclusion of this pattern. By combining relevant contextual information with specific instructions, this pattern addresses scenarios where AI needs to adapt to dynamic inputs or tasks, such as code reviews or architecture recommendations.
    \item \textbf{Question Pattern} - Existing studies \cite{PPforLLM,FewShotLearners} emphasize how targeted question prompts enhance user interaction and problem-solving capabilities. The question pattern is particularly suitable for developers seeking clarifications or detailed explanations from AI assistants.
\end{itemize}

These seven patterns emerged as the most common and impactful in all studies due to their proven ability to enhance task precision, user engagement, and adaptability in various applications. For this reason along with their relevance to software development tasks, we have decided to include them each in our set of prompt patterns for our experiments. 

\subsection{Predicting Prompt Patterns}
Next, we needed to predict which patterns were used in each conversation because they were not specified within the original features of this dataset. We defined a set of keywords for each prompt pattern that are indicative of that pattern in use (whether or not the developer knowingly intended to apply that pattern). The keywords defined for each prompt pattern are outlined in Table \ref{tab:promptkeyword}. After defining the keywords, a Python script will be used to analyze the “PR” and “Issues” data points. This script analyzes prompt patterns in “PR” and “Issues” data points from a JSON file and writes the results to a CSV file. It also records details like the data’s state (open or closed), time lapsed before it was closed, and the number of prompts. Finally, it saves the detected patterns, along with additional metadata, to a CSV file for further review. Table \ref{tab:startsforsample} illustrates the statistics for the sample dataset.

\begin{table}[]
\caption{Specific keywords used for identifying each Prompt Pattern}
\label{tab:promptkeyword}
\resizebox{\columnwidth}{!}{%
{\small
\begin{tabular}{lll} 
\hline
\textbf{Pattern} & \textbf{Keywords} \\
\hline
Persona Pattern & “you are”, “act as”, “pretend to be”, “pretend you are”\\
\hline
Recipe Pattern & “step-by-step”, “recipe”, “guide”\\
\hline
Template Pattern & “template”, “formatting”
 \\
\hline
Output Automator Pattern & “script”, “code”, “executable” \\
\hline
Instructions-Based Pattern & “explain”, “describe”, “list”, “tell me”, “give me”\\
\hline
Context and Instructions Pattern & “based on”, “with this information”\\
\hline
Question Pattern & “what”, “where”, “when”, “who”, “why”\\
\hline
\end{tabular}%
}
}
\end{table}

\subsection{Grouping of Patterns for Evaluation}
Then, once each data point (conversation) was classified based on the prompt pattern used, we were able to begin evaluation. To evaluate the prompt patterns used as a whole, we grouped the conversations according to which prompt pattern was used. Then, we had groups of conversations for each prompt pattern we were evaluating.

\subsection{Ranking Prompt Patterns}
Finally, we evaluate the efficiency of the prompt patterns based on the average number of conversation iterations between the developer and the AI assistant required to elicit a satisfactory response. Prompt patterns will then be ranked in order of most to least efficient, indicating which patterns are the best for use by developers seeking to leverage AI assistants in their coding endeavors. 
Based on the results, we can suggest which prompt patterns are the best for use by developers based on their overall ability to elicit a satisfactory response in the least amount of iterations.

\begin{table}[]
\caption{Statistics for sample dataset}
\label{tab:startsforsample}
{\small
\begin{tabular}{l|l|l}
\hline
\multirow{2}{*}{\textbf{Metrics}} & \multicolumn{2}{l}{\textbf{Values}} \\ \cline{2-3} 
 & \multicolumn{1}{l|}{\textbf{PR}} & \textbf{Issues} \\ \hline
Total Pull Requests & 413 & 250 \\
Total Number of Prompts & 1730 & 1818 \\
Average Number of Prompts per Pull request & 4.19 & 3.85 \\
Open Pull Requests & 387 & 253 \\
Closed Pull Requests & 26 & 219 \\ \hline
\end{tabular}%
}
\end{table}

\section{Experimental Results}
\label{Section:Result}


\subsection{Experimental Setup}
\label{exp-setup}

Our dataset consists of conversations exported from the public DevGPT GitHub repository, containing various conversations between developers and ChatGPT as an AI assistant. 
The DevGPT dataset presented various sets of developer-AI conversations, from which we decided to narrow it down to only two for our experiments. From all options within the DevGPT dataset, we chose to use PR and Issues only because each dataset indicates when the PR or issue was closed/open \cite{SDgithub}. This meant we could determine how effective the prompt type was because we had a metric for how many iterations of conversation between the developer and AI assistant were conducted prior. 

\begin{table}[h]
\centering
\caption{Data before and after data preprocessing}
\label{tab:DataComparison}
{\small
\begin{tabular}{l|ll}
\hline
\multicolumn{1}{c|}{\multirow{2}{*}{\textbf{Metrics}}} & \multicolumn{2}{c}{\textbf{Number of Records}} \\ \cline{2-3} 
\multicolumn{1}{c}{} & \multicolumn{1}{|c|}{\textbf{PR}} & \multicolumn{1}{c}{\textbf{Issues}} \\ \hline
Total & \multicolumn{1}{l|}{7976} & 12618 \\ 
With Patterns & \multicolumn{1}{l|}{3827} & 12,618 \\ 
With State = Open & \multicolumn{1}{l|}{312} & 7357 \\ 
With State = Closed & \multicolumn{1}{l|}{3515} & 5261 \\ \hline
\end{tabular}%
}
\end{table}

Python is the primary tool for analyzing the DevGPT data. It utilizes simple libraries such as JSON for importing the dataset from a JSON file, CSV for exporting detected patterns into a CSV format, and DateTime for handling date and time calculations. The analysis focuses on identifying various prompt engineering techniques by searching for specific keywords associated with different patterns. The script processes the data by iterating through the conversations, extracting relevant information like the state of the conversation (open or closed) and the time elapsed between creation and closure. The detection of prompt patterns is based on the body of the conversations. The results are then compiled and exported into a CSV file, providing a structured summary of the detected patterns, their corresponding conversation states, and additional metadata like the number of prompts and URLs for further reference. Table \ref{tab:DataComparison} shows the data before and after preprocessing.

\subsection{RQ1: \textit{What prompt patterns can be used to elicit a satisfactory response from ChatGPT in less than 5 interactions between the human developer and the AI assistant?}}
In exploring prompt patterns, we initially reviewed both open and closed states in the dataset, which includes pull requests (PRs) and issues. However, determining the efficiency of prompt patterns in open states proved difficult, as these PRs and issues were unresolved, making it unclear whether a satisfactory outcome was achieved. For this reason, we shifted our focus to closed states, where we could evaluate efficiency based on the number of interactions needed to reach a satisfactory outcome.

We chose the threshold of 5 prompts based on our initial analysis of the data. We found that the mean number of interactions required to achieve a satisfactory, closed outcome was around 4, calculated specifically from the ``Number of Prompts" column when the ``Status" column was set to ``Closed." Using this average as a benchmark, we determined that setting the threshold at 5 prompts would allow us to capture prompt patterns that closely align with the average interaction count while providing some flexibility. This threshold was instrumental in helping us identify patterns that consistently yield successful results with minimal back-and-forth, supporting our goal of optimizing prompt efficiency.

\textit{PR Analysis:} For PR records, we analyzed a subset of 412 entries by filtering for closed states with fewer than 5 prompts. This subset enabled us to evaluate the patterns that were most effective in minimizing interaction. We observed that the Output Automator Pattern appeared 78 times, followed by the Simple Instruction Pattern with 67 instances and the Question Pattern with 66 instances. These patterns were particularly effective in achieving satisfactory outcomes with minimal interactions, providing clear and goal-oriented guidance that allowed developers to reach closure with fewer interactions. These results show that the Output Automator Pattern and Simple Instruction Pattern are particularly effective for PRs, facilitating quick responses and minimal interaction by giving clear, structured prompts.

\textit{Issues Analysis:} Similarly, for issues records, we focused on 471 entries under the same conditions, targeting closed states with fewer than 5 prompts. We found that the Question Pattern appeared 63 times, followed by the Output Automator Pattern with 61 instances and the Simple Instruction Pattern with 33 instances. These patterns were most efficient in resolving issues with fewer interactions, especially where clear questions or straightforward instructions were beneficial in directing ChatGPT toward a satisfactory response. These findings highlight that, for issues, the Question Pattern and Output Automator Pattern were particularly effective, supporting efficient problem resolution through direct questions and structured automation guidance.


\subsection{RQ2: \textit{Which prompt patterns are the most efficient? That is, which patterns can be used by developers to elicit a satisfactory response in the fewest possible interactions?}}
From RQ1 analysis, we identified the Output Automator Pattern, Instructions-Based Pattern, and Question Pattern as the most efficient prompt patterns for developers to elicit a satisfactory response in the fewest possible interactions. To back this observation, we applied the following techniques:
\begin{enumerate}
    \item Applied the prompt pattern label using the specific keywords associated with each prompt pattern as mentioned subsection \ref{exp-setup} on the full dataset which contains 7975 PR records and 12,618 Issue records.
    \item Filtered out the records which has State == ``Open" because the pattern associated with these records cannot help identify the efficient patterns.
    \item Identified the correlation between the columns using a correlation matrix.
    \item Calculated the effectiveness of the prompt.
    \item Split the prompt pattern column and create a new row with the split prompt pattern, if more than one prompt pattern has been identified in conversation.
    \item Finding the correlation between the effectiveness of prompt and other columns.
    \item Lastly, finding the average effectiveness score for each prompt pattern and the average number of prompt required by each pattern.
\end{enumerate}

\begin{figure*}[h]
\centering
    \begin{subfigure}{0.23\textwidth}
        \includegraphics[width=\linewidth]{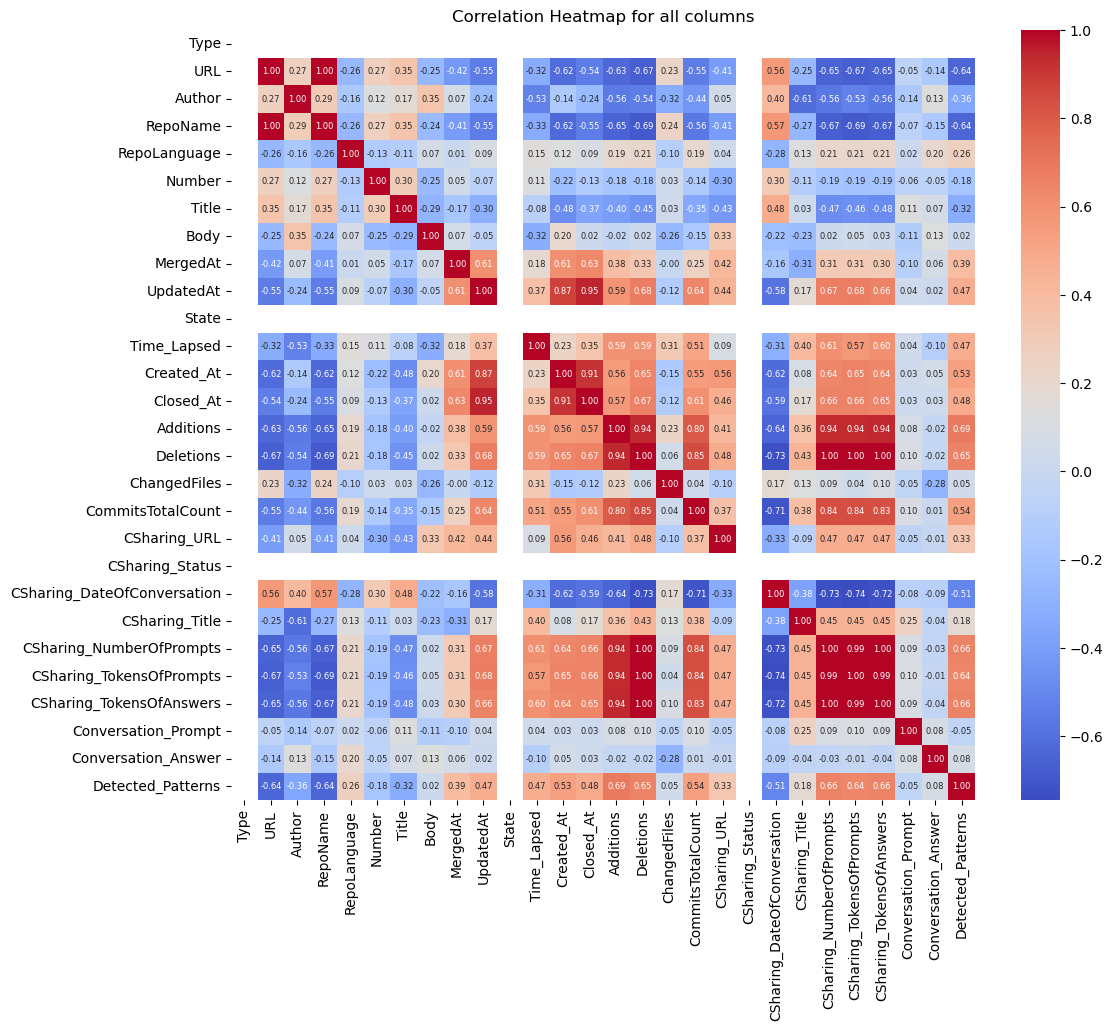}
    \caption{\small{Heatmap before adding Effectiveness Scores for PR dataset}}
    \label{fig:RQ2HeatmapDetect}
    \end{subfigure}\hspace{1em}%
    \begin{subfigure}{0.23\textwidth}
        \includegraphics[width=\linewidth]{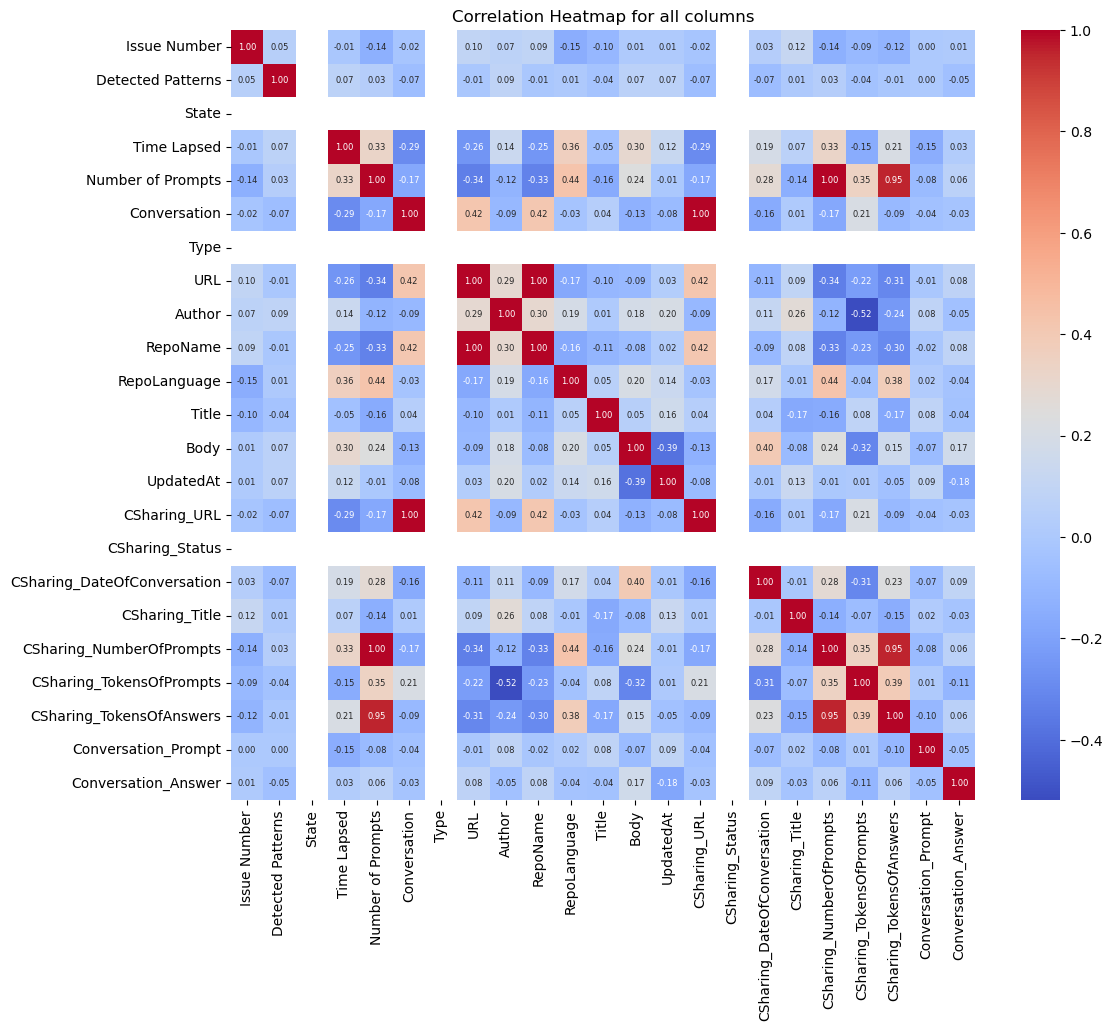}
       \caption{\small{Heatmap before adding Effectiveness Scores for Issues dataset}}
    \label{fig:RQ2IssuesHeatmapDetect}
    \end{subfigure}
    \begin{subfigure}{0.23\textwidth}
        \includegraphics[width=\linewidth]{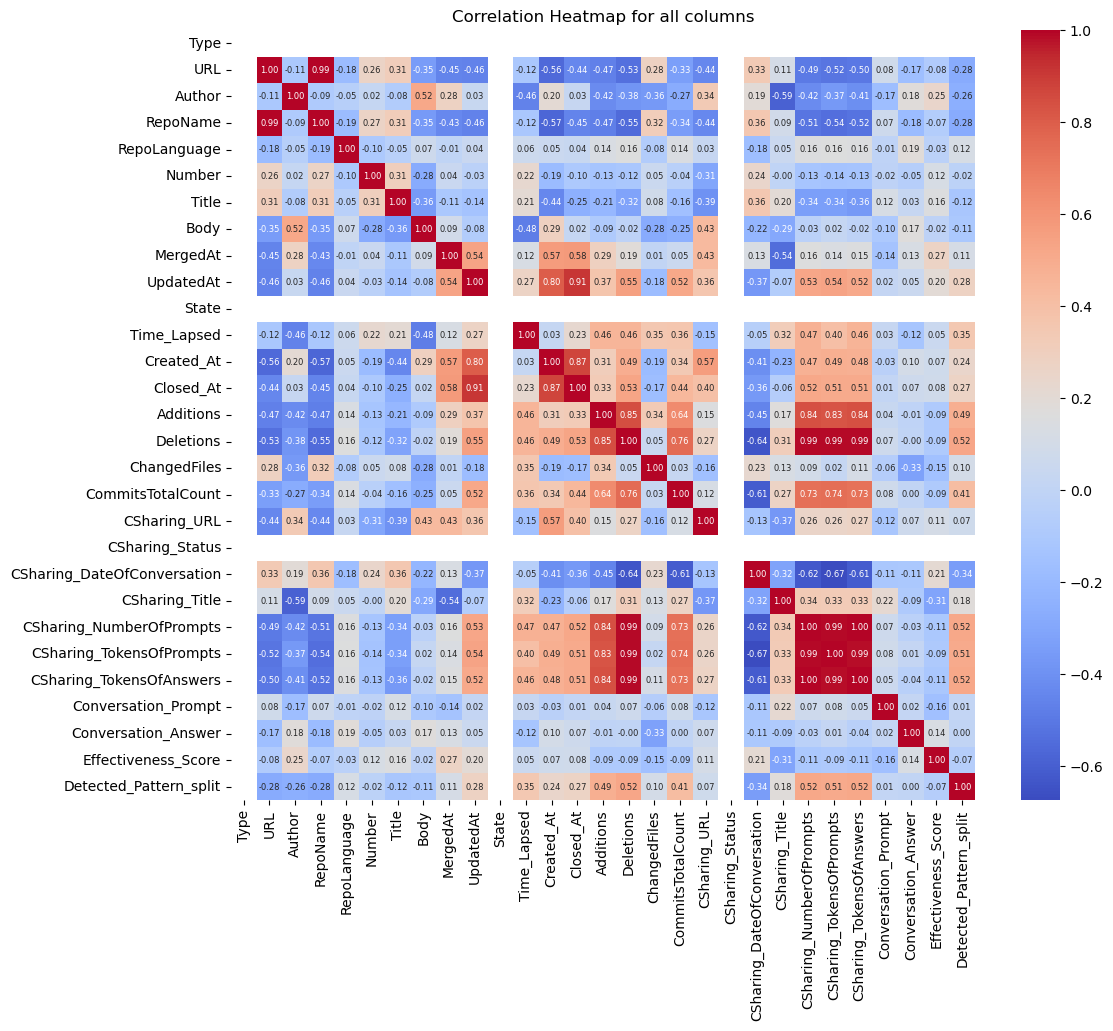}
 \caption{\small{Heatmap after adding Effectiveness Scores for PR dataset}}
    \label{fig:RQ2HeatmapALL}
    \end{subfigure}
    \begin{subfigure}{0.23\textwidth}
        \includegraphics[width=\linewidth]{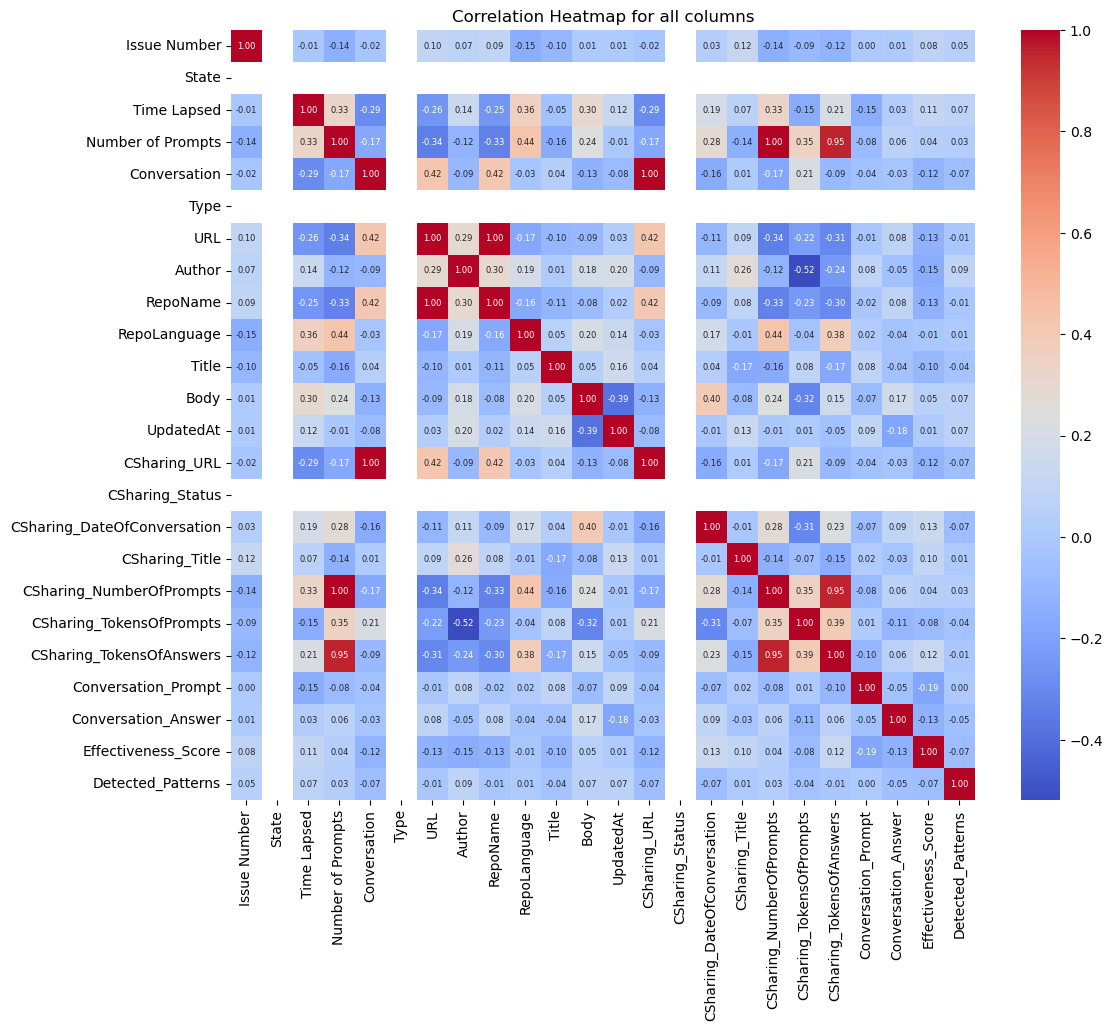}
     \caption{\small{Heatmap after adding Effectiveness Scores for Issues dataset}}
    \label{fig:RQ2IssuesHeatmapALL}
    \end{subfigure}
    \caption{Heatmap before and after adding Effectiveness Scores for PR and Issues}
    \label{fig:parts}

 \end{figure*}

For the first step, we used the same technique for labeling the dataset with prompt patterns as we applied on our subset of RQ1. For the second step, we used State column value and removed the records which had State == ``Open". For further step, we moved forward with records which had State == ``Closed". After applying this filter, the number of records in PR and Issues remained as 3515 and 5261 respectively. The code for all analysis steps can be accessed at \cite{promptpatternsanalysis}. For the third step, we used the heatmap function seaborn python library to find the correlation between the columns for both the PR and Issues dataset (Figures \ref{fig:RQ2HeatmapDetect}\ref{fig:RQ2IssuesHeatmapDetect}). For the fourth step, we preprocessed the data and chose the relevant columns like Conversation\_Prompt, Conversation\_Answer, CSharing\_TokensOfPrompts and CSharing\_TokensOfAnswers for calculating the effectiveness of the prompt. The effectiveness score is calculated as a weighted sum of three components:

\noindent\textbf{Effectiveness of Different Parts}

\begin{align}
\text{Effectiveness Score} = & \, 0.5 \times \text{Response Length Score} \nonumber \\
& + 0.3 \times \text{Prompt-Answer Token Ratio} \nonumber \\
& + 0.2 \times \text{Sentiment Score}
\end{align}

\noindent where:
\[
\text{Response Length Score} = \text{Length of Conversation Answer (in words)}
\]
\[
\text{Prompt-Answer Token Ratio} = \frac{\text{CSharing Tokens of Answers}}{\text{CSharing Tokens of Prompts} + 1}
\]
(Note: The denominator is increased by 1 to avoid division by zero.)
\[
\text{Sentiment Score} = \text{Sentiment polarity of Conversation Answer}
\]
Thus, the final formula for calculating the effectiveness score becomes:
\begin{align}
\textbf{Effectiveness Score} = & \, 0.5 \times \text{Length of Conversation Answer} \nonumber \\
& + 0.3 \times \frac{\text{CSharing\_Tokens\_of\_Answers}}{\text{CSharing\_Tokens\_of\_Prompts} + 1} \nonumber \\
& + 0.2 \times \text{Sentiment polarity of Conversation Ans}
\end{align}


In this function, we calculated the:
\begin{itemize}
    \item Response Length Score: which is the count of words in the Conversation\_Answer column.
    \item Prompt-Answer Token Ratio: this calculates the ratio of token of answer and token of prompt.
    \item Sentiment Score: For this we used, Textblob library which analyzed the sentiment of the Conversation\_Answer column.
    \item Weighted Score: For combining all these scores and the ratio into a single score that gave us an overall quality score of the prompt.
\end{itemize}

Further, we again used the heatmap function (Figures \ref{fig:RQ2HeatmapALL}, \ref{fig:RQ2IssuesHeatmapALL}) to find the correlation between effectiveness\_score and existing columns. After analyzing this heatmap we were able to come up with the features with the effectiveness\_score as target variable \ref{fig:RQ2HeatmapALL}, \ref{fig:RQ2IssuesHeatmapALL}. For a better understanding of the dataset, the prompt patterns column is split into individual patterns by exploding the semicolon-separated and comma-separated values into multiple rows.

To check the presence of an association between the prompt pattern column and the effectiveness score, we used the ANOVA test. The results revealed a statistically significant difference in the effectiveness scores across different patterns, with an F-statistic of 27.04 and a p-value of $4.05 \times 10^{-32}$ for the PR dataset and F-statistic of 41.31 and a p-value of $1.80 \times 10^{-49}$ for Issues dataset. This suggests that the Patterns have a significant impact on the effectiveness scores, indicating varying levels of effectiveness associated with different prompt patterns.
\begin{figure}[h]
    \includegraphics[width=\columnwidth]{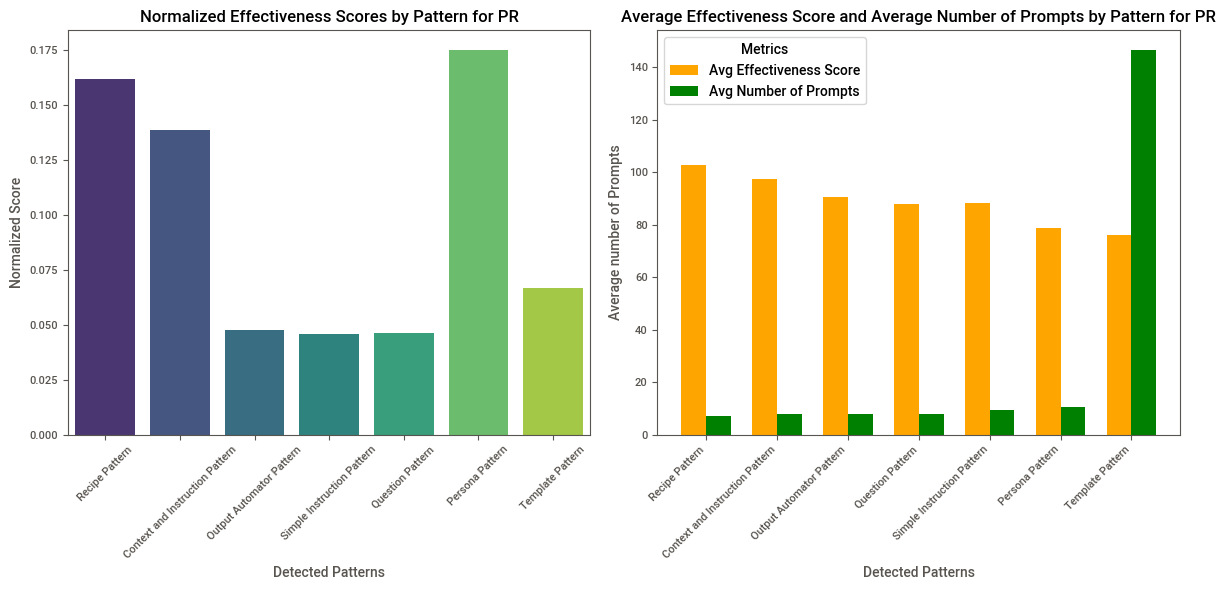}
    \caption{\small{Prompt Patterns and the associated Effectiveness Score for PR}}
    \label{fig:RQ2PR}
\end{figure}
\begin{figure}[h]
    \includegraphics[width=\columnwidth]{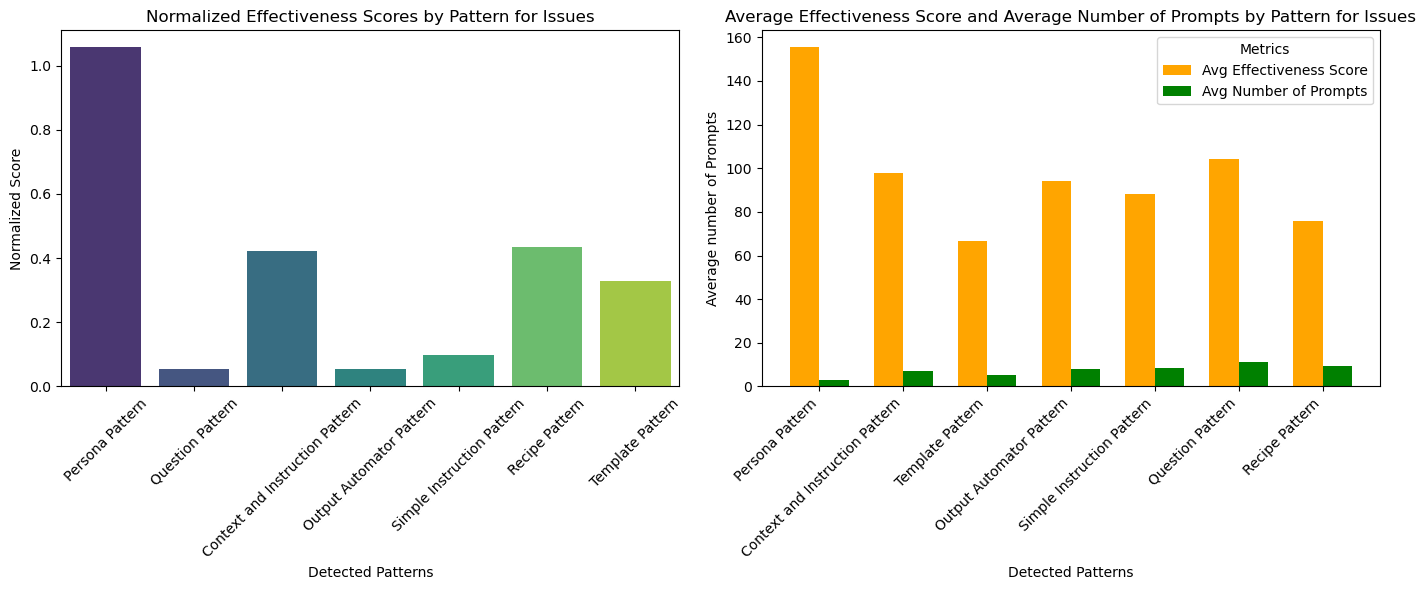}
    \caption{\small{Prompt Patterns and the associated Effectiveness Score for Issues}}
    \label{fig:RQ2Issues}
\end{figure}
Further, we used the effectiveness score for each pattern which is then averaged, and the patterns are ranked based on their mean effectiveness score. Additionally, the frequency of each pattern is calculated, and a Normalized Score is derived by dividing the effectiveness score by the pattern frequency. This normalization allows for better comparison across patterns of varying frequencies (Figures \ref{fig:RQ2PR} and \ref{fig:RQ2Issues}).
Further, we performed a deeper analysis by grouping the data based on the detected Patterns, calculating the average effectiveness score, and the average number of prompts associated with each pattern. A combined metric, Score Ratio, is calculated to identify patterns that exhibit high effectiveness with fewer prompts, indicating more efficient interaction patterns. The patterns are then sorted based on this combined score, providing a ranking of patterns that are both highly effective and efficient in terms of prompt usage. This analysis helps identify the most impactful patterns and their relationship with the number of prompts needed for generating effective responses (Figures \ref{fig:RQ2PR} and \ref{fig:RQ2Issues}).

\begin{table*}[]
\centering
\caption{Statistics related to Prompt Patterns}
\label{tab:promptStats}
\resizebox{\textwidth}{!}{%
\begin{tabular}{lllllllll}
\hline
\multicolumn{1}{c|}{\multirow{2}{*}{\textbf{Patterns}}} & \multicolumn{4}{c|}{\textbf{PR}} & \multicolumn{4}{c}{\textbf{Issues}} \\ \cline{2-9} 
\multicolumn{1}{c|}{} & \multicolumn{1}{l|}{\textbf{Avg. Effectiveness Score}} & \multicolumn{1}{l|}{\textbf{Avg. Number of Prompts}} & \multicolumn{1}{l|}{\textbf{Score Ratio}} & \multicolumn{1}{l|}{\textbf{Frequency}} & \multicolumn{1}{l|}{\textbf{Avg. Effectiveness Score}} & \multicolumn{1}{l|}{\textbf{Avg. Number of Prompts}} & \multicolumn{1}{l|}{\textbf{Score Ratio}} & \multicolumn{1}{l}{\textbf{Frequency}} \\ \hline
\multicolumn{1}{l|}{Persona} & \multicolumn{1}{l|}{78.761773} & \multicolumn{1}{l|}{10.711111} & \multicolumn{1}{l|}{7.353278} & \multicolumn{1}{l|}{450} & \multicolumn{1}{l|}{\textbf{155.414968}} & \multicolumn{1}{l|}{2.938776} & \multicolumn{1}{l|}{\textbf{52.884260}} & \multicolumn{1}{l}{147} \\ 
\multicolumn{1}{l|}{Template} & \multicolumn{1}{l|}{75.942499} & \multicolumn{1}{l|}{\textbf{146.674541}} & \multicolumn{1}{l|}{0.517762} & \multicolumn{1}{l|}{1143} & \multicolumn{1}{l|}{66.779299} & \multicolumn{1}{l|}{5.433498} & \multicolumn{1}{l|}{12.290297} & \multicolumn{1}{l}{203} \\ 
\multicolumn{1}{l|}{Question} & \multicolumn{1}{l|}{87.760302} & \multicolumn{1}{l|}{8.015798} & \multicolumn{1}{l|}{10.948418} & \multicolumn{1}{l|}{1899} & \multicolumn{1}{l|}{104.384771} & \multicolumn{1}{l|}{\textbf{11.156905}} & \multicolumn{1}{l|}{9.356069} & \multicolumn{1}{l}{\textbf{1861}} \\ 
\multicolumn{1}{l|}{Output Automator} & \multicolumn{1}{l|}{90.398429} & \multicolumn{1}{l|}{7.989474} & \multicolumn{1}{l|}{11.314691} & \multicolumn{1}{l|}{1900} & \multicolumn{1}{l|}{94.211404} & \multicolumn{1}{l|}{8.220068} & \multicolumn{1}{l|}{11.461146} & \multicolumn{1}{l}{1754} \\ 
\multicolumn{1}{l|}{Recipe} & \multicolumn{1}{l|}{\textbf{102.644441}} & \multicolumn{1}{l|}{7.072441} & \multicolumn{1}{l|}{\textbf{14.513298}} & \multicolumn{1}{l|}{635} & \multicolumn{1}{l|}{75.838237} & \multicolumn{1}{l|}{9.360000} & \multicolumn{1}{l|}{8.102376} & \multicolumn{1}{l}{175} \\ 
\multicolumn{1}{l|}{Simple Instruction} & \multicolumn{1}{l|}{88.351193} & \multicolumn{1}{l|}{9.242361} & \multicolumn{1}{l|}{9.559374} & \multicolumn{1}{l|}{\textbf{1931}} & \multicolumn{1}{l|}{88.156714} & \multicolumn{1}{l|}{8.703371} & \multicolumn{1}{l|}{10.129031} & \multicolumn{1}{l}{890} \\ 
\multicolumn{1}{l|}{Context and Instruction} & \multicolumn{1}{l|}{97.334039} & \multicolumn{1}{l|}{8.042674} & \multicolumn{1}{l|}{12.102198} & \multicolumn{1}{l|}{703} & \multicolumn{1}{l|}{97.889631} & \multicolumn{1}{l|}{7.207792} & \multicolumn{1}{l|}{13.581084} & \multicolumn{1}{l}{231} \\ \hline
\end{tabular}%
}
\end{table*}

The results from the above analysis are mentioned in Table \ref{tab:promptStats}. From table \ref{tab:promptStats}, we can conclude that: (1) The \textbf{``Context and Instruction"} pattern emerged as the most efficient across both the PR and Issues datasets. In the PR data, it achieved an average effectiveness score of 97.33 with 8.04 prompts, yielding the highest score ratio of 12.10. Similarly, in the Issues dataset, it achieved a score of 97.88 with 7.21 prompts, resulting in a score ratio of 13.58, demonstrating its reliability across various contexts, (2) In the PR dataset, the \textbf{``Recipe"} pattern exhibited the highest average effectiveness score of 102.64 and the lowest average number of prompts (7.07), with a score ratio of 14.51, making it ideal for structured tasks. In the Issues dataset, the \textbf{``Template"} pattern required only 5.43 prompts on average, yielding a score ratio of 12.29, indicating its efficiency for straightforward tasks, and (3) The \textbf{``Output Automator"} pattern in PR showed balanced effectiveness (90.39) and prompt efficiency (7.99 prompts), with a score ratio of 11.31. The \textbf{``Question"} pattern in Issues had the highest effectiveness score (104.38) but required more prompts (11.16), resulting in a lower score ratio (9.35).


\subsection{RQ3: \textit{How does the most efficient pattern compare to the simple Question prompt pattern (the most commonly used prompt by human developers using an AI assistant)}?}

This research was motivated by the inefficiencies developers face with AI-assisted code generation. Prior studies and the initial findings in our research pointed to the challenges of using simple, unstructured prompts, which often lead to iterative and time-consuming exchanges between developers and AI tools.

The findings in RQ2 demonstrate that structured patterns, like ``Context and Instruction," significantly outperform simple question-based prompts. This superiority stems from their ability to integrate context with clear instructions, reducing ambiguity and guiding the AI more effectively.

\noindent\textbf{Structured Patterns: An Empirical Advantage.}

\begin{itemize}
    \item \textbf{Context and Instruction}: This pattern showed consistent reliability, as evidenced by its high effectiveness scores in both the PR and Issues datasets. By providing detailed instructions alongside necessary context, developers could minimize back-and-forth iterations. This approach aligns with findings from earlier studies in our literature review, emphasizing that well-crafted prompts improve AI comprehension and output quality.
    \item \textbf{Task-Specific Patterns}: Patterns such as ``Recipe" and "Template" excelled in scenarios requiring clarity or repetitive formatting. The experimental results suggest that by matching prompt patterns to task requirements, developers can further reduce the interaction count, achieving precise outcomes faster.
\end{itemize}

\noindent\textbf{Why Simple Questions Fall Short?}
Despite their intuitive appeal, simple questions often lack the structure necessary to guide AI effectively. The data showed that simple question-based prompts required more iterations, with a higher likelihood of clarification requests. This aligns with previous research indicating that unstructured prompts lead to suboptimal outcomes due to the AI's need for additional context to generate accurate answers.

\section{Conclusion}
\label{Section:Conclusion}
To conclude, this study demonstrates the significant impact of structured prompt patterns on optimizing interactions between developers and AI-based tools like ChatGPT. By employing patterns such as ``Context and Instruction," ``Recipe," and ``Output Automator," developers can achieve more precise outputs in fewer iterations, ultimately saving time and improving workflow efficiency. The analysis underscores the value of thoughtful prompt engineering as a means to enhance developer productivity and elevate the utility of AI in software development. Future research could expand on these findings by testing prompt patterns across different AI models and software development contexts, further refining best practices for maximizing AI's potential in coding assistance.





\bibliographystyle{abbrv}
\bibliography{sample-base}

\end{document}